\documentclass[twocolumn,showpacs,preprintnumbers,amsmath,amssymb]{revtex4}
\usepackage[english]{babel}
\usepackage[latin2]{inputenc}
\usepackage{epsfig}

\begin{document}

\title{Lattice Expansion of (Ga,Mn)As:\\
The Role of Substitutional Mn and of the Compensating Defects}

\author{\sc J.~Ma\v{s}ek and F.~M\'{a}ca}

\address{Institute of Physics, AS CR, Na Slovance 2, 182 21 Prague
8, Czech Republic}


\begin{abstract}
We apply the density-functional technique to determine the lattice
constant of GaAs supercells containing Mn$_{\rm Ga}$, Mn$_{\rm
int}$, and As$_{\rm Ga}$ impurities, and use a linear
interpolation to describe the dependence of the lattice constant
$a$ of Ga$_{1-x}$Mn$_{x}$As on the concentrations of these
impurities. The results of the supercell calculations confirm that
Mn$_{\rm Ga}$ does not contribute to the lattice expansion. The
increase of $a$ is due to both Mn$_{\rm int}$ and As$_{\rm Ga}$,
that are both created in the as-grown (Ga,Mn)As in proportion to
$x$, and that are most probably present in a remarkable amount
also in the best annealed materials.
\end{abstract}

\maketitle

\section{Introduction}

Diluted magnetic semiconductors (DMS), represented by (Ga,Mn)As
mixed crystals, are particularly interesting materials because of
their hole-mediated ferromagnetism \cite{Ohno98, Dietl00}. This,
combined with the semiconducting behavior, makes the DMS
attractive for applications in spin electronics, and stimulated
extensive theoretical and experimental studies of these materials
in the last years.

Recently, the compositional dependence of the lattice constant of
the (Ga,Mn)As mixed crystals has been investigated with respect to
Mn incorporation in the lattice. It is known for long that the
lattice constant of (Ga,Mn)As increases with increasing content of
Mn \cite{Ohno99}. Originally, it was attributed to the
substitution of Mn for Ga in the cation sublattice and it was
assumed that the lattice constant of  (Ga,Mn)As extrapolates to
the lattice constant of the zinc-blende MnAs crystal that is
larger than the lattice constant of GaAs \cite{Schott03}. This
assumption is, however, in a contradiction with the simple
estimate based on the atomic radii of the constituent atoms
because Mn atom is smaller (0.117 nm) than Ga atom (0.125 nm)
\cite{Langes73}. Also the density-functional study of ${\rm
\alpha}$-MnAs showed \cite{Zhao02} that the lattice constant of
MnAs with a perfect zinc-blende structure is smaller then the
lattice constant of GaAs.

Most recently, the density-functional method of determining the
lattice constant by minimizing the total energy has been applied
also to the (Ga,Mn)As. The mixed crystals with realistic
concentrations of Mn up to 10 percent were treated by means of the
LMTO method within the coherent-potential approximation (CPA)
\cite{Masek03}. It was found that the substitutional Mn (Mn$_{\rm
Ga}$) has only a negligible effect on the lattice constant in this
concentration range. It was deduced that the observed increase of
the lattice parameter is a secondary effect caused by creation of
an increasing number of compensating donors, proportionally to the
content of Mn. In particular, the As antisite defects (As$_{\rm
Ga}$)  and Mn atoms in the interstitial positions (Mn$_{\rm int}$)
were shown to cause a remarkable increase of the lattice constant
of (Ga,Mn)As. Quantitatively, the dependence of the lattice
parameter $a(x_s, x_i,y)$ on the partial concentrations of
Mn$_{\rm Ga}$ Ga, Mn$_{\rm int}$, and As$_{\rm Ga}$ was
parametrized in the form
\begin{equation}
a(x_s,x_i,y) = a_{o} + 0.002 x_s + 0.105 x_i + 0.069 y {\rm
~~(nm)},
\end{equation}
where $a_{o}$ is the lattice constant of the pure GaAs.

In strongly compensated materials it is expected that the number
of Mn$_{\rm int}$ and As$_{\rm Ga}$ increases proportionally to
the total concentration $x = x_s + x_i$ of Mn \cite{Masek04b}. The
dependence of $a$ on $x$ can be simply estimated from Eq. (1) in
the limiting case of the complete compensation, i.e. for $x_s = 2
x_i + 2 y$. In this case, we obtain a linear dependence $a(x)
\approx a_{o} + 0.035 x$~(nm). The coefficient will be smaller for
partial compensation which is in a reasonable agreement with the
experimental value 0.032 nm \cite{Ohno99}.

The theoretical prediction that the increase of the lattice
parameter $a$ with Mn concentration is, at least partly, connected
with the presence of the interstitial Mn, was also confirmed by
recent measurements \cite{Kuryliszyn04,Sadowski04,Zhao04} in which
the lattice constants for as-grown and annealed materials were
compared. During the the post-growth thermal treatment, the
interstitial Mn atoms diffuse out of the material and the lattice
constant decreases accordingly. These measurements, however,
differ in several respects from the calculations. First of all,
the calculated increase of the lattice constant due to Mn$_{\rm
int}$ according to Eq. (1) seems overestimated by a factor of 2.
Second, even the best materials with minimum compensation (i.e.
with all interstitials removed) have their lattice constant larger
than GaAs and increasing with the Mn content. It is interpretted
that - after all - the Mn substitution itself can expand the
lattice.

That is why we re-examine our original coherent-potential study
\cite{Masek03} by using the full-potential
linearized-augmented-plane-wave method (FPLAPW \cite{WIEN}) that
overcomes some simplifications involved in the LMTO-CPA study
(atomic-sphere approximation, unrelaxed lattice, etc.). We apply
it to the supercells of GaAs with Mn$_{\rm Ga}$, Mn$_{\rm int}$,
and As$_{\rm Ga}$ impurities.

\section{Results}

We use tetragonal supercells formed by 8 and 16 molecular units of
GaAs. These supercells, containing a single defect (Mn$_{\rm Ga}$,
Mn$_{\rm int}$, or As$_{\rm Ga}$), represent materials with 12.5
and 6.25 atomic percent of the impurities, respectively. The
lattice constant was determined by minimizing the total energy
$E_{tot}$ of the supercell with respect to $a$, either with or
without lattice relaxation. In practice, the dependence of
$E_{tot}$ on $a$ was approximated by a cubic polynomial fitted in
approx. 10 points.

\begin{figure}
\begin{center}
\includegraphics[width=53mm,height=85mm,angle=270]{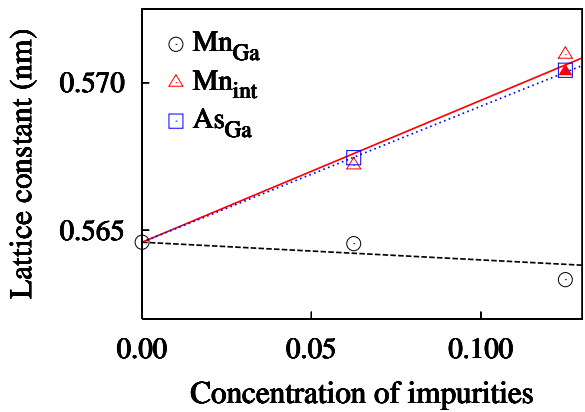}

\caption{Calculated lattice constant as a function of the
concentration of the impurities: (a) Mn atoms in the
substitutional positions (circles), (b) Mn atoms in the
interstitial positions (triangles), (c) As antisite defects
(boxes). The full triangle corresponds to a relaxed geometry.}
\end{center}
\end{figure}

The results are summarized in Fig. 1. First of all, the
calculations confirm that the substitutional Mn$_{\rm Ga}$ atoms
do not contribute to the expansion of the lattice. The dependence
of the lattice constant on $x_{s}$ is weak and, in contrast to the
CPA results, decreasing. Assuming that the lattice constant of
(Ga,Mn)As follows the Vegard's law, i.e. that the function
$a(x_{s})$ is linear in the entire concentration range, we obtain
the linear coefficient $\approx$~-0.005~nm. This value is in a
reasonable agreement with the interpolation between the lattice
constants obtained for GaAs and ${\rm \alpha}$-MnAs \cite{Zhao02}.

The interstitial Mn atoms were considered in both tetrahedral
positions, T(As$_{4}$) and T(Ga$_{4}$). Although the local
relaxation of the lattice is different for these two defects \cite
{Masek04a}, the effect on the lattice expansion is almost the same
for both. That is why only the data for T(As$_{4}$) are presented
in Fig. 1. Without relaxation (empty triangles), the lattice
expands significantly in presence of Mn$_{\rm int}$ and the
lattice constant increases almost linearly with $x_{i}$. If the
relaxation around  Mn$_{\rm int}$ is taken into account, the
expansion is slightly weaker, as indicated by a full triangle in
Fig. 1. It is important to notice that the linear coefficient is
$\approx$~0.048~nm now, approximately one half of the value
obtained from the CPA calculations.

The contribution of the As$_{\rm Ga}$ antisite defects to the
lattice expansion is very similar to Mn$_{\rm int}$. Assuming that
the contributions of various defects to the lattice expansion are
additive, we summarize the above results into a simple linear
formula,
\begin{equation}
a(x_s,x_i,y) = a_{o} - 0.005 x_s + 0.048 x_i + 0.046 y {\rm
~~(nm)}.
\end{equation}

\section{Discussion}

Eq. (2) is an full-potential counterpart of Eq. (1) that was
obtained from the TB-LMTO-CPA study \cite{Masek03}. Because of
several simplifications involved in the CPA calculations
(effective medium, mimimum basis, atomic-sphere approximation,
etc.) Eq. (2) should be considered more reliable. At the same
time, however, the sensitivity of the parameters of Eqs. (1,2) to
the method of calculations indicates that also the coefficients in
Eq. (2) represents only a rough quantitative estimate for the
compositional dependence of the lattice constant and should be
used with a caution.

\begin{figure}[b]
\begin{center}
\includegraphics[width=53mm,height=85mm,angle=270]{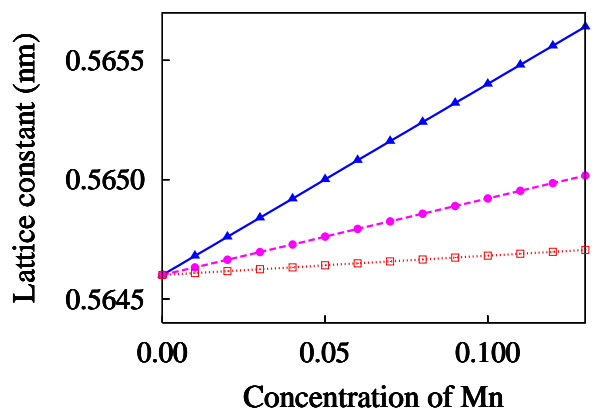}

\caption{Concentration dependent lattice constant of (Ga,Mn)As
according to Eqs. (2] and (4). (a) as-grown material with Mn$_{\rm
Ga}$, Mn$_{\rm int}$, and As$_{\rm Ga}$ (triangles), (b) mixed
crystal with all Mn$_{\rm int}$ removed (boxes), (c) material with
the $x_{i}$ reduced to one third (circles).}
\end{center}
\end{figure}

The most important features of Eq. (2) is that it confirms the
recent results \cite{Zhao02,Masek03} that the lattice expansion of
(Ga,Mn)As can not be attributed to the substitutional Mn. It is
related to the presence of the interstitial Mn and As$_{\rm Ga}$
antisite defects as well. The contribution of these two intrinsic
defects to the expansion is characterized by a linear coefficient
$\approx$~0.05~nm instead of much larger values resulting from the
CPA calculations. The present results are much closer to the
experiment \cite{Kuryliszyn04,Sadowski04,Zhao04}.

To obtain a more detailed picture of the lattice expansion of real
materials and to compare the role of Mn$_{\rm int}$ and As$_{\rm
Ga}$ we need to estimate the partial concentrations of Mn$_{\rm
Ga}$, Mn$_{\rm int}$, and As$_{\rm Ga}$. To do this, we start with
the concentration dependent formation energies $E({\rm Mn}_{\rm
Ga})$, $E({\rm Mn}_{\rm int})$, and $E({\rm As}_{\rm Ga})$
\cite{Masek02,Masek04b} and assume that the corresponding partial
concentrations in the as-grown (Ga,Mn)As can be approximately
calculated from the dynamical-equilibrium conditions
\cite{Masek04b,Masek05}
\begin{equation}
E({\rm Mn}_{\rm Ga}) = E({\rm Mn}_{\rm int}), \hspace{3mm} E({\rm
As}_{\rm Ga}) = 0 .
\end{equation}
Using for simplicity the linearized form of the formation energies
we obtain that, except for the lowest values of $x$, the partial
concentrations of substitutional and interstitial Mn, and also the
concentration of the As$_{\rm Ga}$ antisite defects increase
proportionally to $x$,
\begin{equation}
x_{s} \approx 0.85 x, \hspace{3mm} x_{i} \approx 0.15 x,
\hspace{3mm} y \approx 0.11 x.
\end{equation}

Combining Eq. (2) with the estimate Eq. (4), we arrive to the
theoretical model of the concentration dependence of the lattice
constant in the as-grown (Ga,Mn)As. It is shown by solid line in
Fig. 2. The dotted and dashed lines in Fig. 2 correspond to the
annealed materials with a reduced amount of the interstitial Mn,
but with the unchanged concentration of the more stable Mn$_{\rm
Ga}$ and As$_{\rm Ga}$. If all Mn$_{\rm int}$ are removed, the
lattice constant is still an increasing function of $x$, but its
slope is much smaller. An intermediate case with the number of the
Mn interstitials reduced to one third is considered as a realistic
example. Most probably, the observed increase of the lattice
constant in the annealed materials
\cite{Kuryliszyn04,Sadowski04,Zhao04} indicates that a large
number of the compensating antisite defects (As$_{\rm Ga}$ as well
as some residual Mn$_{\rm int}$) are present even in the best
annealed samples with the minimum compensation.

To summarize, the full-potential supercell calculations confirmed
qualitatively the recent results of the CPA studies \cite{Masek03}
of the concentration dependence of (Ga,Mn)As mixed crystals. The
present results are closer to the experiment. They show the
negligible influence of the substitutional Mn on the lattice
constant and the role of the intrinsic compensating donors in the
lattice expansion.\\

\subsection* {Acknowledgment} This work has been done
within the project AVOZ1-010-0520 of the AS CR. The financial
support was provided by the Academy of Sciences of the Czech
Republic (Grant No. A1010214) and by the Grant Agency of the Czech
Republic (202/04/583).

\end{document}